\documentclass[conference]{IEEEtran}
\IEEEoverridecommandlockouts
\usepackage{cite}
\usepackage{amsmath,amssymb,amsfonts}
\usepackage{algorithmic}
\usepackage{graphicx}
\usepackage{subfigure}
\usepackage{textcomp}
\usepackage[table]{xcolor}
\usepackage{multirow}
\usepackage{array}
\usepackage{slashbox}
\usepackage{url}
\def\BibTeX{{\rm B\kern-.05em{\sc i\kern-.025em b}\kern-.08em
    T\kern-.1667em\lower.7ex\hbox{E}\kern-.125emX}}

\begin{document}

\title{Understanding Rowhammer Attacks through the Lens of a Unified Reference Framework}

\author{
\IEEEauthorblockN{Xiaoxuan Lou\IEEEauthorrefmark{1},  Fan Zhang\IEEEauthorrefmark{1}, Zheng Leong Chua\IEEEauthorrefmark{2}, Zhenkai Liang\IEEEauthorrefmark{2}, Yueqiang Cheng\IEEEauthorrefmark{3},Yajin Zhou\IEEEauthorrefmark{1}}

\IEEEauthorblockA{\IEEEauthorrefmark{1}Zhejiang University \IEEEauthorrefmark{2}National University of Singapore£¬ \IEEEauthorrefmark{3}Baidu XLab}

\IEEEauthorblockA{\{xiaoxuanlou, fanzhang, yajin\_zhou\}@zju.edu.cn, \{chuazl, liangzk\}@comp.nus.edu.sg, chengyueqiang@baidu.com}
}

\maketitle

\begin{abstract}
Rowhammer is a hardware-based bug that allows the attacker to modify the data in the memory without accessing it, just repeatedly and frequently accessing (or hammering) physically adjacent memory rows. So that it can break the memory isolation between processes, which is seen as the cornerstone of modern system security, exposing the sensitive data to unauthorized and imperceptible corruption. A number of previous works have leveraged the rowhammer bug to achieve various critical attacks.

In this work, we propose a unified reference framework for analyzing the rowhammer attacks, indicating three necessary factors in a practical rowhammer attack: the \emph{attack origin}, the \emph{intended implication} and the \emph{methodology}. Each factor includes multiple primitives, the attacker can select primitives from three factors to constitute an effective attack. In particular, the methodology further summarizes all existing attack techniques, that are used to achieve its three primitives: \emph{Location Preparation (LP)}, \emph{Rapid Hammering (RH)}, and \emph{Exploit Verification (EV)}. Based on the reference framework, we analyze all previous rowhammer attacks and corresponding countermeasures. Our analysis shows that how primitives in different factors are combined and used in previous attacks, and thus points out new possibility of rowhammer attacks, enabling proactive prevention before it causes harm. Under the framework, we propose a novel {\em expressive} rowhammer attack that is capable of accumulating injected memory changes and achieving rich attack semantics. We conclude by outlining future research directions.
\end{abstract}

\section{Introduction} \label{sec_intro}
In order to meet the growing demand of memory capacity in the modern computer system, the cell density of DRAM module keeps increasing. However, such high density aggravates
electromagnetic interference between memory cells, which can eventually corrupt the data stored in the memory, \textit{i.e.}, causing the affected bit to {\em flip}. The attacker can enhance such interference to induce bit
flip on purpose by accessing the memory in controllable patterns, such as repeatedly and frequently accessing, often called {\em hammering}. This attack is commonly known as \emph{rowhammer}.
The nature of rowhammer attacks is to use software code to induce the faults into underlying hardware in a controllable manner, which is difficult to prevent.
Since Kim et al.~\cite{kimfirstpaper} identified the rowhammer bug and demonstrated its pervasiveness in the modern DRAM modules in 2014, the number and variety of attacks leveraging such bug have been steadily increasing.
As the rowhammer attack allows the attacker to alter the data in the memory without directly accessing it, the isolation between different
processes, that is seen as the cornerstone of system security and
previously thought to be safe, is severely affected. Therefore, such bug not only influences memory reliability but also causes serious security breaches.

The rowhammer attack has been actively explored by researchers,
where new applications have been constantly proposed in various environments.
For instance, the attacker can corrupt page tables to gain kernel privileges from user space
processes in the Linux~\cite{linuxkernel} or android system~\cite{drammer,guardion}, while
they can also modify shared binary files in the memory to achieve privilege escalation~\cite{anotherflip}.
On the virtual machine platform, such as a public cloud, the attacker is able to break the memory isolation between the VM
and the host~\cite{xiaoyuancloud}, and access other co-host VMs
without authorization~\cite{flipfengshui}.
As for the remote rowhammer attacks, which are usually originated from the website or network, the attacker can trigger
faults on remote hardware with
JavaScript from a malicious website~\cite{rowhammerjs}, gain arbitrary memory read and
write accesses in the browser~\cite{dedupmachina} or even subvert the remote system only by sending network packets~\cite{nethammer,throwhammer}.

To better understand the landscape of rowhammer attacks, there is a need to organize the various rowhammer attacks in order to better analyze them.
To this end, we propose a {\em unified reference framework} to interpret the similarities and differences
of components in rowhammer attacks. The understanding will be a useful
guide for researchers to identify new attacks and defences.
In its most basic form, a rowhammer attack is the manifestation of a bit flip due to repeatedly accessed memory.
This can be broken down into three
primitives: \emph{Location Preparation (LP)}, \emph{Rapid Hammering (RH)}, and
\emph{Exploit Verification (EV)}.
In each primitive, we identify common abstractions used from existing work, providing us with a set of techniques which can achieve the intended primitive.
These primitives form the core of our analysis, {\em methodology} of a rowhammer attacker.

Though the hammering methodology has much in common, different attack techniques have different ways (or {\em origins}) to drive the repeated access and to use the changed memory (or {\em implications}) which serves as constraints on the available primitives that can be utilized.
Therefore, a practical attack includes three necessary factors: the {\em attack origin}, the {\em intended implication} and the {\em methodology}.
Using these observations, we designed a unified reference framework which is used to analyze the composition of existing rowhammer attacks and corresponding countermeasures.

Though our analysis, we also discuss limitation of primitives in past
attacks, so that researchers may find unworked combination of primitives or
even detect new primitives to achieve novel rowhammer attacks.
Through our analysis, we add a new primitive \emph{Store Error (SE)} to the methodology and identify a novel rowhammer attack that is
able to accumulate memory changes and achieve expressiveness of attack semantics.

In summary, we made the following contributions in this paper.
\begin{itemize}
  \item Presenting a unified analysis framework, that indicates three factors for understanding rowhammer attacks.
  \item Summarizing all existing attack techniques and providing the methodology for conducting a rowhammer attack.
  \item Analyzing and classifying existing rowhammer attacks and countermeasures under the unified framework.
  \item Proposing a novel expressive rowhammer attack based on the framework and also discussing other possible future directions of rowhammer research.
\end{itemize}

\section{Background} \label{sec_bg}
In this section, we provide necessary background for this paper, showing the DRAM's architecture and recent development, also introducing the principle of rowhammer bug.

\subsection{DRAM}
\begin{figure}[h]
  \centering
  \includegraphics[scale=0.3]{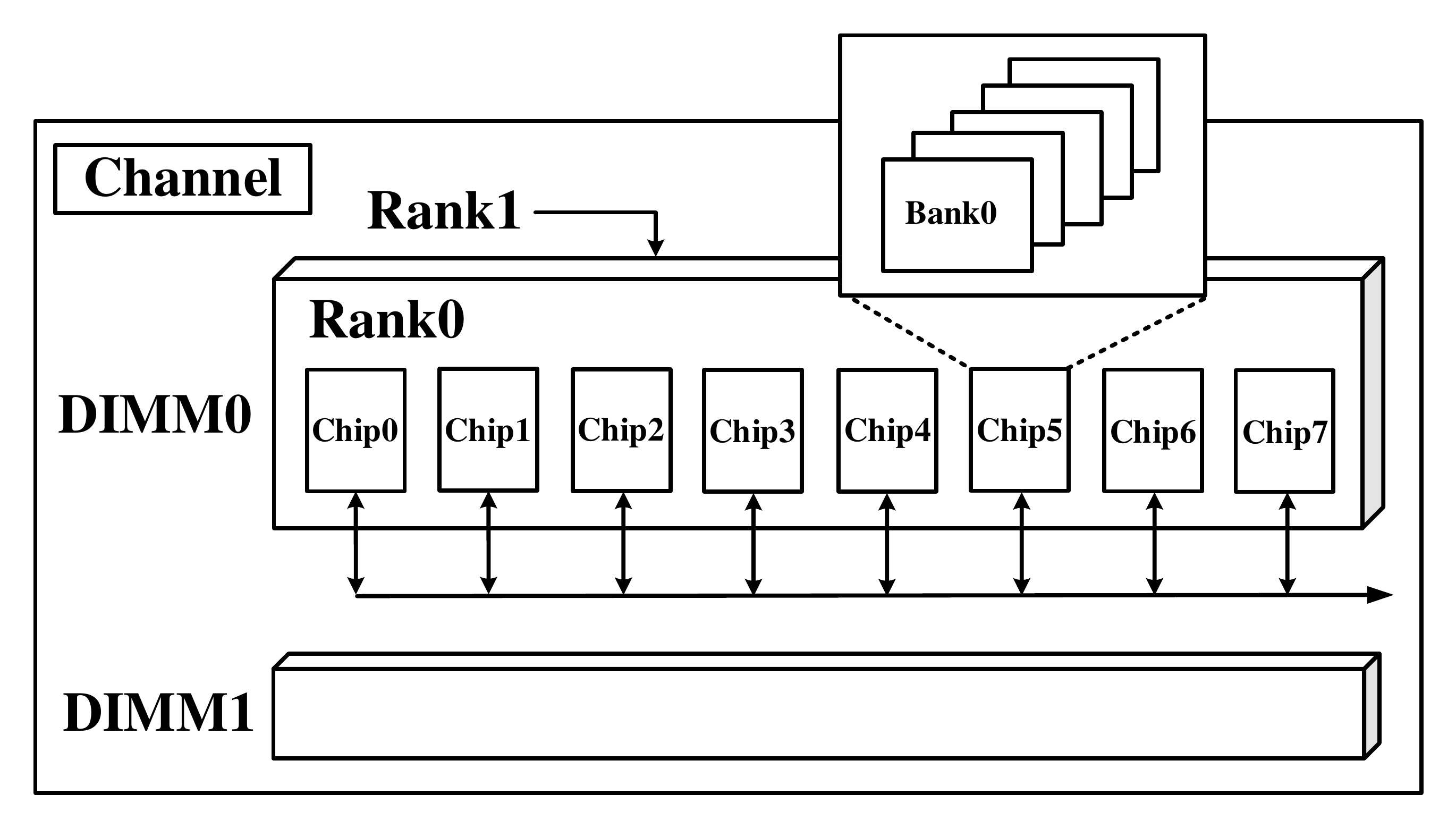}\\
  \caption{DRAM architecture}\label{fig_DRAM}
\end{figure}

As the most commonly used memory module, DRAM (\emph{Dynamic Random Access Memory}) has much simpler structure and larger capacity than other memory modules such as SRAM (\emph{Static Random Access Memory}). In modern memory systems, the DRAM modules need to be encapsulated and assembled before they can be connected to the CPU, hence there are multiple layers between CPU and DRAM modules. First the memory
system is generally organized in multiple memory channels and each of them is handled by dedicated memory controller. Every channel contains
multiple DIMMs (\emph{Dual Inline Memory Module}), \textit{i.e.}, the physical modules on the motherboard. The DIMM consists of one or two ranks, which
normally are the sides of the physical module. A rank is further partitioned into multiple DRAM chips or multiple banks. The bank is an array
of memory rows and has a row buffer to cache the latest accessed row. The typical architecture of DRAM is illustrated in Fig.~\ref{fig_DRAM}.

The DRAM cell is made up of a transistor and a capacitor. The transistor controls the access to the cell while the capacitor stores the bit value by being charged or discharged. As charges that are held in the DRAM decays, in order to ensure the integrity of the data, DRAM cells are required to be periodically refreshed. The refresh interval should not be longer than 64 ms, typically between 32 and 64 ms.
Increased demand of memory capacity in recent years resulted in the growing density of DRAM cells. As a result, electrical interference between cells have become more severe.
This electrical interference between memory cells is much easier to affect the charge state of the capacitor inducing bit flips known as \emph{disturbance errors}~\cite{kimfirstpaper}.

\subsection{Rowhammer Bug}
The occurrence of rowhammer is an unanticipated result during the improvement process of DRAM modules. Modern high-density DRAM is threatened by the potential disturbance errors, and the rowhammer bug deliberately amplifies the threat. It activates a memory region in special patterns to exacerbate charge leaking of cells in that area, corrupting the sensitive data stored in the memory. In practical attacks, the special pattern used for rowhammer is usually to repeatedly access certain memory rows with a high frequency, just similar as ``hammering'' memory rows. That is why such bug is called as \emph{rowhammer}~\cite{kimfirstpaper}. The assembly code \texttt{Attack$\_$Loop} used on Intel/AMD machines for inducing rowhammer bug is shown below. The instruction \texttt{clflush} can flush data from the cache to ensure the access indeed reaches to the DRAM.

~\\
\fbox{\shortstack[l]{\texttt{\small 1~Attack$\_$Loop:} \\ \texttt{\small 2~~~mov~~(addr$\_$X),~$\%$rax}~~// \texttt{read the row X}~ \\ \texttt{\small 3~~~mov~~(addr$\_$Y),~$\%$rbx}~~// \texttt{read the row Y}~ \\ \texttt{\small 4~~~clflush~~(addr$\_$X)}~~// \texttt{flush X from cache}~ \\ \texttt{\small 5~~~clflush~~(addr$\_$Y)}~~// \texttt{flush Y from cache}~ \\ \texttt{\small 6~~~jmp~~Attack$\_$Loop} }}
~\\

\noindent Due to the existence of row buffer, two aggressor rows \texttt{X} and \texttt{Y} are alternately opened and closed in the attack code, in order to avoid reading data from row buffer rather than DRAM.
By hammering aggressor rows, the attacker can induce bit flips in the adjacent memory region, and then further exploit it to compromise system.

Nowadays, the vulnerability of almost all existing DRAM modules under rowhammer bug has already been carefully assessed, including DDR3~\cite{kimfirstpaper,aichinger2014known,park2014active} and
DDR4~\cite{micheletti2013tuning,ddrerror,lanteigne2016rowhammer,aga2017good}.
The assessment results show that both DDR3 and DDR4 are vulnerable from the rowhammer bug.
Meanwhile, a number of patents aiming to leverage the rowhammer bug have been filed~\cite{bains2015method,bains2015row,bains2016row,bains2016distributed,greenfield2015row,greenfield2014method}.
Recently, some rowhammer-like attacks are also exploited on other memory modules such as Flash memory \cite{nandflash,woots,krautter2018fpgahammer}, and their principle are similar with rowhammer attacks on the DRAM.
As a widespread and threatening security issue, the rowhammer bug deserves to be analyzed carefully, so that we further propose a unified reference framework for understanding these powerful attacks leveraging the rowhammer bug.

\section{Unified Reference Framework}
\label{sec_model}

\subsection{Overview of the Unified Reference Framework}
The framework is designed based on the observations we made when investigating existing rowhammer attacks. To conduct a practical rowhammer attack, first where the attack originates from and what are the affects of the attack need to be confirmed. Then both of them can determine the actual techniques that are used to perform the attack. As such, the framework can be broken down into three factors: the attack origin, the intended implication and the methodology.

\noindent \textbf{Attack Origin.}
The attack origin is the site where attacker locates. In different sites, the attacker owns different privileges which greatly affect the types of techniques he can use. This factor allows the analyst to better predict the possible attack surface that is available to the attacker and the ease of performing a specific rowhammer attack by identifying just the threat origin. Examples of possible attack origins are local processes, website, network, etc. The further classification of attack origin is introduced in Section~\ref{sec_origin}.

\noindent \textbf{Intended Implication.}
The intended implication describes the intended result of a rowhammer attack, \textit{i.e.}, the corruption of selected attack target. Similar to the attack origin, this factor also affects the design of actual attack method, since different types of targets require different attack techniques to be used for corruption. Besides that, it also allows for better prediction over the possible attack vectors. We will describe the possible intended implications in Section~\ref{sec_implication}.

\noindent \textbf{Methodology.}
The methodology factor in the framework provides a systematic way of organizing and analyzing the different technical approaches used in a rowhammer attack. To achieve this, we provide a description of the life-cycle of a rowhammer attack and the corresponding techniques that are required to achieve a specific primitive in the life-cycle. Details are discussed in Section~\ref{sec_methodology}.

\begin{figure}
  \centering
  \includegraphics[scale=0.15]{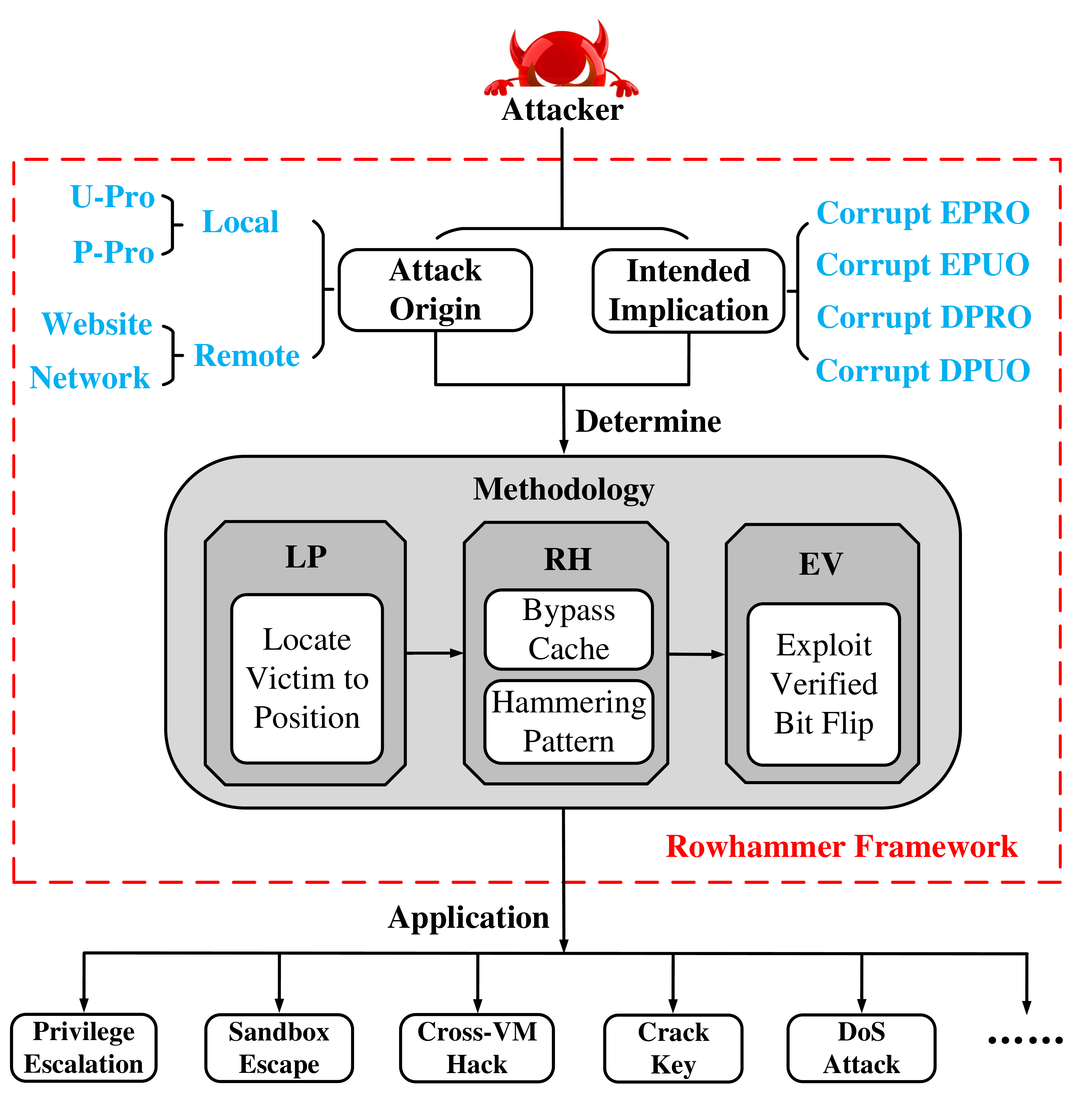}\\
  \caption{Unified framework of rowhammer attacks.}\label{fig_framework}
\end{figure}

The three factors above make up the unified analysis framework. Each factor contains multiple primitives, and all presented rowhammer attacks can be analyzed based on primitives extracted from three factors.
Fig.~\ref{fig_framework} illustrates the framework of rowhammer attacks. The attacker can leverage the framework to understand the available attack surface and limitations of an attack, maximizing effectiveness resulting in various practical applications, such as privilege escalation, sandbox escape, cross-VM hack and so on.

\subsection{Attack Origin}\label{sec_origin}
In general, the origins of rowhammer attacks can be classified to either that of a local or remote origin.
Every origin has its own
privilege set in the system. For example, the local attack origin has the capabilities to access native resources, abuse system interfaces, or even actively utilize code to create opportunities to manipulate the memory. The remote origin usually has more constraints and only has limited attack surface to the victim system. For example, a remote attacker is normally not allowed to leverage the instruction \texttt{clflush} to flush the cache for performing rapid access on the memory while a typical local attacker is able to. We will discuss these two attack origins in detail below.

The local origin is able to be further differentiated based on the amount of privileges the origin has, in other words, if the attacker is able to execute privileged operations or not. Hence, we further term the local origin as Privileged Process~\emph{(P-Pro)} and Unprivileged Process~\emph{(U-Pro)}. Given the prevalance of the principle of least privilege, most of local rowhammer attack origins are \emph{U-Pro}. For such attack origin, attack techniques which require system permissions are not available. However, other techniques that can be performed in user space are still valid. Typical instances include exhausting the memory to locate the
victim~\cite{linuxkernel,linuxkernelppt, drammer, guardion, stillhammer} or performing hammering pattern which requires no knowledge of address mapping~\cite{linuxkernel,anotherflip}. These attack techniques will be introduced in the methodology.
As for the \emph{P-Pro}, in fact the attacker is impossible to own the full privileges of the victim system, otherwise he doesn't have to attack. But the attacker is able to take a process from some special scenarios, like the VM or the old kernel, to obtain partial privileges. With the help of \emph{P-Pro}, the rowhammer attack has much fewer constraints. It is able to gather in-depth information about the layout of the memory~\cite{DRAMA,jung2016reverse,xiaoyuancloud}, or even control the location of the victim object in memory~\cite{flipfengshui,xiaoyuancloud}.

Compared to the local origin, remote origin is typically restricted in the way it can interact with the victim system, since the remote attacker can only impact the system through the service providing the remote connection. However, in certain scenarios and with the help of specific mechanisms, the attacker can also perform effective rowhammer attack from remote. All presented remote rowhammer attacks are launched from two remote origins, the website and the network. The website is the main remote origin for the rowhammer attacks. Attackers on the website usually use the JavaScript scripts attached in the website to corrupt the victim's sensitive data, trying to escape from the sandbox and affect the host~\cite{rowhammerjs,dedupmachina}. The achievement of website attacks greatly increase the threat and practicability of rowhammer.
Beyond that, two recent works propose the remote rowhammer attacks originated from the network~\cite{throwhammer,nethammer}.
Previously, the general assumption is that the network is not fast enough to provide a high-frequency packet transmission that is required for the rapid hammering to induce bit flips.
This changed with the advent of ever faster network speeds making rowhammer attacks over the network possible.
It has been demonstrated that with 500Mbps~\cite{nethammer} and 10Gbps~\cite{throwhammer} network, the remote rowhammer attack can exploit the flipped bit to impact the targeted system in practical ways.

Note that this classification for origins is not fixed and can be updated according to the new scenarios if possible. The importance lies in the intuition and methodology behind the classification. More details about how these origins leverage techniques from methodology to conduct attack will be introduced in Section~\ref{sec_attacks}.

\subsection{Intended Implication}\label{sec_implication}
The effect of a rowhammer attack is to induce a value change into the targeted object which is \textbf{(1)} in a different memory partition from the attacker and \textbf{(2)} the attacker himself has no write permission to it.
Hence, all intended implications of rowhammer attacks can be classified based on the type of attack target, using the privilege level of the memory partition where the target is located and if its read permission is granted to the attacker or not. With that, the targets of rowhammer attacks can be categorized  to four classes:
\begin{description}
	\item[(a)]\emph{Equal Privileged Readable Object (EPRO)}
	\item[(b)]\emph{Equal Privileged Unreadable Object (EPUO)}
	\item[(c)]\emph{Different Privileged Readable Object (DPRO)}
	\item[(d)]\emph{Different Privileged Unreadable Object (DPUO)}
\end{description}
Moreover, the intended implication of a rowhammer attack is to corrupt one of above four targeted objects.

\noindent \textbf{Corrupt EPRO.}
In this case, the attacker and the victim are of the same privilege level, such as two user processes, and the attacker is capable of reading the context of the victim object. For instance, in the work of \cite{linuxkernel}, they achieve the sandbox escape
by corrupting the instruction flow in the NaCl sandbox, which is at the equal privilege level of the
attacker and allows attacker to read the modification in the code. On the cloud, the VM owned by the attacker may share some particular objects with other equal privileged VMs, so that the attacker can corrupt these objects and read their content to verify whether there are any bit flips~\cite{flipfengshui,xiaoyuancloud}.

\noindent \textbf{Corrupt EPUO.}
Normally, due to the isolation between different processes, one process cannot read the value of another equal privileged one without system permissions. Hence, the \emph{Equal Privileged Unreadable Object (EPUO)} is much commoner than \emph{EPRO}.
For example, the pointers in a process, which cannot be directly read by another process, is able to be corrupted and used for further exploiting~\cite{dedupmachina,throwhammer}.
The SGX enclave is also not allowed to be directly accessed, and the attacker can induce a
bit flip in it and trigger the halt of system, resulting in a DoS attack~\cite{jang2017sgx,nethammer}.

\noindent \textbf{Corrupt DPRO.}
Both of the two aforementioned classes corrupt objects that hold the equal privilege with the attacker. However, to make the attacks more powerful and threatening, the attacker generally
aims to corrupt those objects which have different, or more precisely, higher privileges, in order to elevate its own privilege.
For example, the shared library
is a typical \emph{Different Privileged Readable Object (DPRO)}, the attacker is able to corrupt opcode in the .so file to acquire higher system privilege~\cite{anotherflip}.

\noindent \textbf{Corrupt DPUO.}
In surveyed rowhammer literatures, \emph{Different Privileged Unreadable Object (DPUO)} is the most common attack target, as it is usually closely related to the authorization and system security. The most typical DPUO is the page table. Since it controls the address mapping of the process, the page table is always taken as the first target of attacks aiming to privilege escalations. A large number of presented rowhammer works demonstrate how to corrupt the page table and gain higher privilege, such as gaining kernel privilege on the Linux~\cite{linuxkernel}, leveraging JavaScript to subvert system~\cite{rowhammerjs}, obtaining the root privilege on mobile~\cite{drammer}, etc.

\subsection{Methodology}
\label{sec_methodology}

The process of a rowhammer attack can be divided as following steps. The attacker first
selects suitable vulnerable memory positions for locating the targeted object. After the appropriate setup, the attacker then proceeds to perform
the hammering on the DRAM in order to produce the bit flip. Verification will then be performed to ensure that
the intended exploitable bit error is indeed registered.  As can be seen from the above description, the rowhammer attack life-cycle contains
three primitives: Location Preparation (LP), Rapid Hammering (RH) and Exploit Verification (EV).

\subsubsection{Location Preparation (LP)}
In a DRAM module, the positions of those bits vulnerable from rowhammer are usually fixed. If the attacker wants to impact the system with bit errors caused by rowhammer, first he needs to locate the victim object to a memory position. The object can be modified with desired flipped bits violating its integrity with potential security implications.
For most of rowhammer attacks, locating victim is required as they have to corrupt some specific bits in an object to achieve the attack.
That said, in some special scenarios, the locating of victim is not necessary, like that of crashing the system or DOS attack as they only require random bits to be flipped to achieve the desired effect.
Hence such attacks can directly start from \emph{Rapid Hammering} (RH).
Note that the profiling of entire memory to gain the knowledge of vulnerable bit positions in advance is not the prerequisite for the locating stage.
This is because it can also be achieved by spraying objects or trying different positions. 
There are a number of techniques for locating victim in prior rowhammer works, including:

\noindent\textbf{(A1) Object Spraying.} This technique is to spray a large amount of victim objects (\textit{e.g.}, page table) to nearly fill the memory, so that with more possibilities the victim is located on the vulnerable memory position where contains intended bit flip~\cite{linuxkernel,linuxkernelppt,rowhammerjs}.

\noindent\textbf{(A2) Forced Padding.} This technique leverages certain system mechanisms, such as buddy allocator, to craft special memory allocation pattern (taking up whole available memory space), then force the OS to pad the victim object to the memory position that is deliberately left~\cite{dedupmachina,drammer,glitch,throwhammer,stillhammer}.

\noindent\textbf{(A3) Induced Replacement.} This technique is to utilize special mechanisms, like memory deduplication or paravirtualization, inducing the OS to replace the victim object with a forged object,
which is counterfeited by the attacker and has the same content with the victim~\cite{xiaoyuancloud,flipfengshui}.

\noindent\textbf{(A4) Try and Abort.} This technique keeps trying different memory positions and aborting undesired attempts, until it finds the vulnerable one whose distribution of flipped bits meets the attack requirements~\cite{anotherflip,pfa}.

\subsubsection{Rapid Hammering (RH)}
As the most pivotal primitive in a rowhammer attack, \emph{Rapid Hammering} injects bit flip to the victim object, which is the key purpose of rowhammer.
To achieve it, the aggressor rows on the DRAM need to be accessed rapidly, which requires attacker to bypass the cache and get to the memory directly.
Once rapid access to memory is achieved, specific memory access patterns can be utilized to maximize the effects of the disturbance and hence the probability of a bit flip.
Details of each step in the hammering process will be discussed below.

\emph{Bypass Cache.} The nature of rowhammer is the charge leaking caused by frequently accessing on the DRAM, but the existing of cache which stores recently accessed data blocks the attacker's effective access to the DRAM. So bypassing cache is necessary for rowhammer attacks and it can be implemented with:

\noindent\textbf{(Ba1) Specific Instructions.} Some system instructions can be called from the user space to flush the cache, such as the most widely used one \texttt{clflush}~\cite{kimfirstpaper}.
Intel also came out the \texttt{clflushopt}, another version of \texttt{clflush} in its Skylake microarchitecture. Besides that, some non-temporal access instructions are also available for fast uncached accesses to the DRAM~\cite{newapproach}.

\noindent\textbf{(Ba2) Cache Eviction Set.} Since cache is organized in multiple slices whose mapping from
physical address is fixed, the access to those addresses mapped to the same cache slice can flush the cache. The set of those congruent addresses is named \emph{cache eviction set}~\cite{rowhammerjs,dedupmachina,glitch}, which can be leveraged for fast accesses to the DRAM.

\noindent\textbf{(Ba3) Uncached memory.} There are some memory regions that can be accessed without going through the cache, such as DMA (Direct Memory Access) memory~\cite{drammer} or RDMA (Remote Direct Memory Access) memory~\cite{throwhammer}. These uncached memory regions offer an available bed for rapid access of rowhammer attacks.

\emph{Hammering Pattern.} The way of conducting hammering greatly affects the efficiency of flipping bits with rowhammer, so that it should be cautiously selected depending on the actual attack environment.

\begin{figure}
	\centering
	\subfigure[Single-sided]{\includegraphics[scale=0.5]{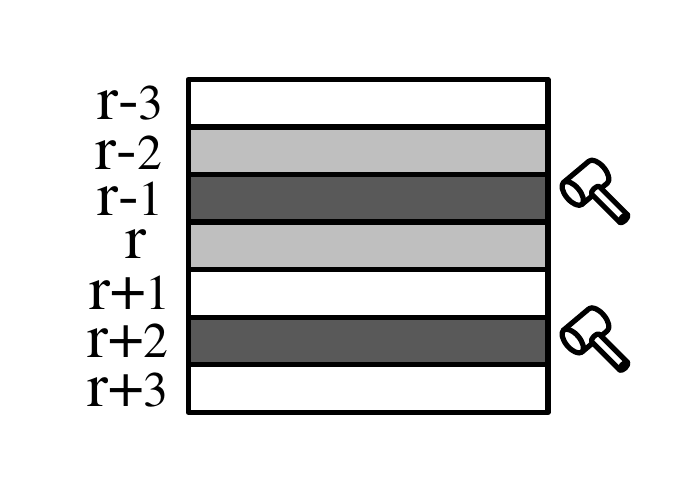}%
		\label{fig_singleside}}
	\hfil
	\subfigure[Double-sided]{\includegraphics[scale=0.5]{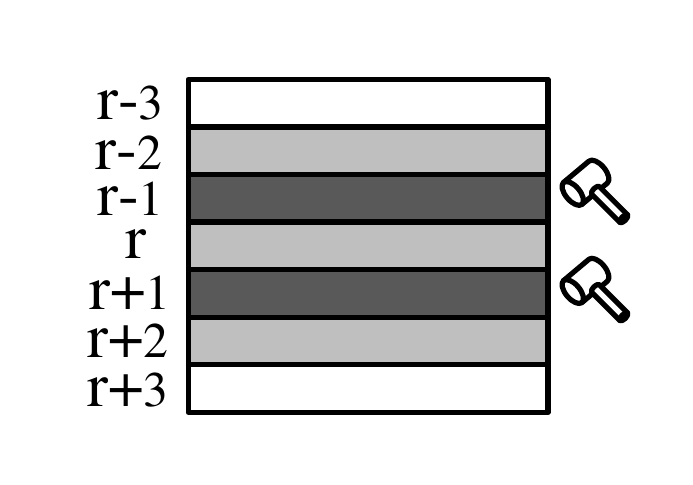}%
		\label{fig_doubleside}}
    \hfil
	\subfigure[One-location]{\includegraphics[scale=0.5]{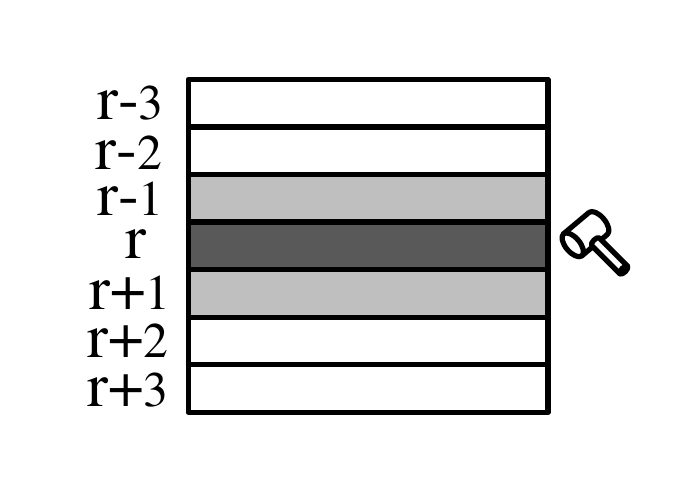}%
		\label{fig_onelocation}}
	\caption{Schematic of three hammering patterns: the black rows marked with a hammer are the aggressor rows and the gray rows are the victim rows.}
	\label{fig_rhpattern}
\end{figure}

\noindent\textbf{(Bb1) Single-sided Hammering.} Take random two memory rows as the aggressor rows and alternately access them, which is possible to induce bit flips in the adjacent rows. As shown in the
Fig.~\ref{fig_singleside}, such pattern only hammers the victim row from single side, which is generally flexible but slow.

\noindent\textbf{(Bb2) Double-sided Hammering.} This pattern marks two rows adjacent to the victim row as the aggressor rows, as shown in the Fig.~\ref{fig_doubleside}. The double-sided hammering
pattern is more efficient but complicated than the single-sided one. In the practical attacks,
such pattern usually requires the knowledge of virtual-to-physical mappings or at least a large physically contiguous memory region. Consequently it normally need to leverage system interface (e.g., \emph{/proc/self/pagemap}) or special mechanisms (\textit{e.g.} huge page).

\noindent\textbf{(Bb3) One-location Hammering.} With the usage of close-page policy or adaptive policy in the system, the hammering to the same row can also induce bit flips in the memory, as the cached row in the row buffer would be automatically evicted by the OS. The Fig.~\ref{fig_onelocation} shows the pattern of one-location hammering, the attacker just needs to hammer one aggressor row for injecting bit errors. Usually such pattern is weaker and slower than the other two patterns, but it is much stealthier as it requires no privileges~\cite{anotherflip,nethammer}.

\subsubsection{Exploit Verification (EV)}
With the primitive \emph{RH}, one possible bit might be flipped in the victim object, however the corresponding error may not be the one that is desired. Hence, the attacker needs to conduct the verification and then he can continue to exploit it for further applications. The verification of injected bit flips is an important stage in the rowhammer attack. Based on the intended implication, \textit{i.e.}, the targeted object corrupted by the attack, the techniques used for verification are different, which can be roughly categorized into two types:

\noindent\textbf{(C1) Directly Read.} If the target is readable for both parties (the attacker and the victim), the attacker can directly read the memory to check if the injected
bit flip is achieved. 

\noindent\textbf{(C2) Behavior Judgment.} Sometimes since the target is not readable, the attacker cannot verify the result through a simple read. Therefore, the verification of injected bit flips has to be inferred by observing the victim's behavior. If the victim behaves as expected, the injected bit flip can be considered as a success.

The exploiting of verified bit flips is the last step to achieve intended implication, which can be further leveraged to generate various practical security applications.
In terms of the privilege level of the target, the attacker can escalate his privilege by corrupting higher privileged objects, or compromise inaccessible data by corrupting equal privileged objects.
The primitive \emph{Exploit Verification} is also the major target of current rowhammer countermeasures, more details about such defence will
be introduced in Section \ref{sec_counter}.

\section{Attacks} \label{sec_attacks}
In this section, we collect all presented rowhammer attacks from previous literatures and analyze them with our unified framework. Based on different attack origins, we demonstrate how to leverage the methodology to design the process of a rowhammer attack to achieve intended implications. We also discuss possible security applications
that can be exploited from the achieved implications.

\subsection{Local Rowhammer Attacks}
Local attack usually is the least constrained one among all types of attacks, since it has the most resources that is available to approach its target. Similarly, the rowhammer attack driven by the local origin is also the most powerful type. 
Hence, a large amount of previous rowhammer attacks aim to be performed locally~\cite{linuxkernel,drammer,guardion,curious,anotherflip,stillhammer,pfa,xiaoyuancloud,flipfengshui}.
As previously described, the local rowhammer attack can be categorized into either Unprivileged Process \emph{(U-Pro)} or Privileged Process \emph{(P-Pro)} dependent on the privilege of the local process.

\noindent\textbf{Unprivileged Process \emph{(U-Pro)}.}
For the security of the system, most user-mode processes are unprivileged, so that they cannot access sensitive data without system permissions. It effectively prevents the malicious user from modifying other processes' data or corrupting kernel. 
However, the presence of rowhammer attack breaks this security mechanism. As rowhammer attack can modify the data without accessing it, even an unprivileged process can cross the domain boundary and corrupt targeted object that was impossible to be touched. Based on the proposed rowhammer framework, in order to design a rowhammer attack that aims to realize certain intend implication,
first the techniques that are available on the current origin need to be selected and then combined to conduct the process of actual attacks.

In the primitive \emph{Location Preparation}, the victim is required to be steered towards the memory position which has the desired bit flip. However, as the \emph{U-Pro} cannot manipulate memory at will, directly allocating the victim to that position is infeasible. This is because the memory allocation is controlled by the MMU and all details of the physical address layer are invisible to normal users. To cope with the problem, multiple novel techniques in the methodology are adopted.
First, an empirical way is to use the technique \emph{(A1) Object Spraying} which exhausts all available memory regions, hoping one of them contains the targeted position~\cite{linuxkernel,linuxkernelppt}. This method is simple, however it causes large memory footprints, which obviously affects the performance and alerts the victim. Therefore currently it is not that preferred. Another probabilistic technique for locating the victim is to use the technique \emph{(A4) Try and Abort}, which tries to test every position to find a desired one~\cite{anotherflip,pfa}. The method is stealthy and hard to be discovered as it does not increase the memory pressure. However since the probability that the victim to be located on the desired position is usually very small, the locating stage takes a long time. Hence it is a trade-off between stealth and speed. Noting that during the locating, the attacker usually leverages side-channel techniques to leak the address information of victim to promote the attacks. Besides these probabilistic techniques, in order to implement deterministic locating,
the \emph{U-Pro} can also leverage those primitives aiming to steer the victim precisely, such as the technique \emph{(A2) Forced Padding}. It can be achieved with the help of certain system mechanisms,
such as the buddy allocator, in the user space. By abusing buddy allocator to reuse and partition memory in a predictable way, the attacker can occupy free memory at a fine-grained level and force the OS to pad the victim to a controlled position. The so-called \emph{Phys Feng Shui} method~\cite{drammer} is a typical example, and its variations are used in many works~\cite{guardion,stillhammer,glitch,throwhammer}.

As for the \emph{Rapid Hammering} on the victim, the attacker needs to (1) bypass the cache and (2) choose the hammering pattern. On the x86 platform, the \emph{U-Pro} is allowed to leverage \emph{(Ba1) Specific Instructions}, like the \texttt{clflush}~\cite{kimfirstpaper}, to flush the cache  ensuring access to the DRAM instead of the cache.
Sometimes such instruction is disabled, like in the NaCl sandbox~\cite{linuxkernel}, the attacker may can utilize other non-temporal instructions to achieve it~\cite{newapproach}.
However in the ARM platform, since the cache flush instructions are privileged and non-temporal instructions are not available, the attacker has to introduce the \emph{(Ba3) Uncached Memory} to perform rapid access. On the mobile devices, the DMA memory becomes the main attack vector, as it can offer uncached access to the memory and open a contiguous physical memory space~\cite{drammer,glitch}. Besides that, the technique \emph{(Ba2) Cache Eviction Set} is also a candidate for an unprivileged attacker who can neither use cache flush instruction nor leverage uncached memory~\cite{rowhammerjs,anvil,curious}.
With the ability of bypassing the cache, the attacker is able to select suitable pattern to perform hammering. The efficiency of error injection in the rowhammer attacks would be greatly affected by the hammering pattern,
normally \emph{(Bb1) Single-sided Hammering} is weaker than \emph{(Bb2) Double-sided Hammering} as it
usually induces less bit flips and takes longer time. Whereas, in the rowhammer attacks originated from the \emph{U-Pro}, the single-sided hammering is more applicable as it
requires no privileges, while the double-sided one usually asks for system permissions or special memory mechanisms like huge page or DMA. Recent work~\cite{anotherflip} proposes another new
hammering pattern named \emph{(Bb3) One-location Hammering}. They claim that both single-sided hammering and double-sided hammering are mainly based on the open-page policy. When the close-page policy or adaptive policy are enabled, one-location hammering can also induce bit flips~\cite{nethammer}. The experiment results show that the one-location hammering is
effective for bit flip injection, but it is weaker than prior two hammering patterns.

Once the bit is flipped, it needs to be verified whether it is the desired one and then to be successfully exploited. The attacker decides the verification method based on his intended implication. If the intended implication aims to corrupt the readable objects, such as \emph{EPRO} or \emph{DPRO}, he is able to \emph{(C1) Directly Read} the target to verify the bit flip. Otherwise, the technique \emph{(C2) Behaviour Judgement} is adopted to check the validity of injected bit flip, like checking the pointed target of the corrupted page table~\cite{linuxkernel,drammer,stillhammer}. Later, the attacker can exploit the flip to result applications that aim to harm the system. Actually, the intended implication has a deterministic influence to the type of exploited application. If it is to corrupt the equal privileged objects, like \emph{EPRO} or \emph{EPUO}, the attacker can leverage the bit error to affect another user process in local, like the sensitive code in the browser~\cite{linuxkernel}. However, if the implication aims to corrupt the different privileged objects, such as the shared library, the attacker is capable of escalating his privileges to gain the whole system~\cite{anotherflip}.

\noindent\textbf{Privileged Process \emph{(P-Pro)}.}
The local rowhammer attacks originated from \emph{P-Pro} has more privileges than those from \emph{U-Pro}. However, in the actual attacks, the \emph{P-Pro} is generally
provided by some especial scenarios, like the VM or old kernel, which also restrain the extensive use of such attacks.
Compared with \emph{U-Pro}, \emph{P-Pro} allows the attacker to leverage nearly all techniques in the methodology to conduct a rowhammer attack.

First, according to the methodology, the victim is required to be located on the right position. All techniques used for \emph{Location Preparation} in the rowhammer attacks driven by the \emph{U-Pro} are also available for   privileged rowhammer attacks. Beyond that, as the attacker has more system permissions, novel privileged techniques are allowed
to be used, such as the \emph{(A3) Induced Replacement}. the \emph{P-Pro} is able to read the content of the target and forge a fake object that contains the same data. Then he can adopt especial system mechanisms like memory deduplication to merge them as COW (\emph{Copy On Write}) pages, replacing the target with attacker's controlled pages to achieve locating~\cite{flipfengshui}. If the \emph{P-Pro} owns higher privilege, it even can directly guide the OS to redirect from the original object to the forged object, which will replace the attack target~\cite{xiaoyuancloud}. Theoretically, such technique can also be adopted by the \emph{U-Pro}, but as its achievement generally requires certain privileges, it is the last choice in the unprivileged rowhammer attacks.

In the \emph{RH} phase, the privileged attacker is permitted to utilize the system calls or interfaces to flip desired bit in the victim. For instance, he can directly use \texttt{clflush}
instruction to bypass the cache and carry out fast access on the DRAM, or leverage \emph{/proc/self/pagemap} to obtain the virtual-to-physical address mapping~\cite{curious}, which greatly reduces the difficulty of the attack. So for the attacker who owns all primitives, he can perform rapid hammering in a very easy way. First he reads the \emph{pagemap} interface to gain information of address; then finds out physical adjacent rows of victim;
finally runs \emph{(Bb2) Double-sided Hammering} to inject a bit flip. However, the attacker with such high privilege is impractical in the actual attacks, indeed he usually only has partial privileges.
Hence, in general the injection of bit flip is achieved by combining unprivileged techniques with certain privileged techniques.

The primitive \emph{EV} in the privileged rowhammer attacks also depends on the intended implication of the attack. Methods for verifying injected bit flip are determined based on whether the corrupted object is readable, like  what \emph{U-Pro} does. Compared with the unprivileged attacks which mostly aim to cross boundary between kernel and user space, privileged attacks try to exploit the injected error to achieve more interesting applications, as its origin is
already privileged. When the attack origin is on the VM, the intended implication is usually to corrupt objects on the cloud, so that the attacker can
cross boundary between himself and host or other VMs~\cite{flipfengshui,xiaoyuancloud}.
Besides, if the privileged origin relies on the old kernel, the exploitation is similar with that in the unprivileged attacks, just being much easier~\cite{pfa,curious}. But due to its less practicality,
there are only few previous works exploiting in the privileged mode with the help of old kernel.

\subsection{Remote Rowhammer Attacks}
The attacker can conduct a rowhammer attack from a remote client by sending certain aggressive data, like browser scripts or network packets, to the victim system to induce bit flips. By and large, the remote attack is more constrained and harder to fulfill than a
local attack, as all it can do is to send aggressive data remotely and receive the feedback from victim. However from the other side, since the remote attack requires less local resources, it becomes more practical, and as the attack is launched remotely, it is also more stealthy. There are two common remote origins, the website and the network, that are leveraged in literatures~\cite{rowhammerjs,dedupmachina,glitch,throwhammer,nethammer}.

\noindent\textbf{Website.}
In all presented remote rowhammer attacks, the website is the mostly used origin because of its good concealment and wild spreading. The attacker generally uses phishing emails or some other ways to trick the victim user to access a malicious website, then injects attack code and steals victim's information.

For the primitive \emph{LP}, as the attacker is also an unprivileged one who cannot leverage any system privileged interfaces, the victim is required to be located on the desired
position without any help of system privileges. Some \emph{LP} techniques used in the local rowhammer attacks are also available for website rowhammer attacks. A remote attacker can
spray pages to cover all possible targeted positions~\cite{rowhammerjs}, store the target to his controlled area where has the profiled flips~\cite{dedupmachina}, craft the special memory
allocation pattern to force a desired padding~\cite{glitch}, or just keep trying until it finds the right position. However, some techniques, like \emph{(A3) Induced Replacement}, usually need privileges
are not available.

After the victim is located, the attacker aims to inject a bit flip to this remote hardware with the primitive \emph{RH}. Compared to that in local, performing hammering on the remote device from the
website is much more difficult. As the hammering operations originated from the website are required to run with the JavaScript, where there is no concept of virtual addresses or pointers and no
access to physical address mappings, the attacker cannot achieve the hammering like
what native code does in the local attacks. First of all, the \emph{(Ba1) Specific Instructions}, such as the \texttt{clflush}, are not available any more in the JavaScript. In addition, the
browser normally does not use \emph{(Ba3) Uncached memory} to store the data, which makes such technique also unavailable. Therefore, the
\emph{(Ba2) Cache Eviction Set} based on CPU~\cite{rowhammerjs,dedupmachina} or GPU~\cite{glitch} are introduced to the website rowhammer attacks.
By accessing congruent addresses belonging to the same eviction set, the remote attacker can efficiently flush the cache and implement fast memory hammering, which requires neither particular instructions nor mechanisms.
As the identification of cache eviction set is a bit difficult, the technique mostly is adopted in the remote attacks, instead of local attacks that have much more choices.

The website attacker usually tends to adopt \emph{(Bb1) Single-sided Hammering} to inject bit flip, because the constrained \emph{(Bb2) Double-sided Hammering} is much harder for the remote attacker. However, it does not mean that the website attacker cannot drive the double-sided hammering. He can achieve it by leveraging special tools or designing crafted mechanism. For instance, the THP (\emph{Transparent Huge Page}) is usually utilized for the double-sided hammering~\cite{rowhammerjs}, as it can offer a large contiguous physical memory to construct a three-rows region for the hammering. In addition, the technique \emph{(A2) Forced Padding} in the locating stage can hold a large amount of memory which may also contains rows adjacent the victim, so that the attacker can run double-sided hammering on these controlled rows~\cite{glitch}. As for the (Bb3) \emph{One-location Hammering}, there are no previous works to conduct remote rowhammer attack from the website. It might be a new direction.

To verify the injected bit flip, the attacker on the website can \emph{(C1) Directly Read} the victim if the target of intended implication is readable. Otherwise he can look into the action of the victim, once the victim has abnormal behaviors what he expect, meaning the injected bit flip is desired. Then the attacker can exploit the flipped bit to cross security domain boundary and make some applications. The boundary for the attacker on the website
is normally the sandbox of browser, which blocks him from the host. Once flipping one bit in the sandbox, the attacker gains arbitrary memory read and write access in the browser, or even escapes the sandbox to access the system without restriction.

\noindent\textbf{Network.}
Different with rowhammer attacks from other origins, the network rowhammer attack runs on the victim system without any code controlled by the attacker. All the attacker does is
just sending network packets to the targeted system. As a result, the network rowhammer attack is considered as the real remote rowhammer attack and it is quite promising in terms of a new and practical attack. Similarly, the rowhammer attack originated from the network can also be designed based on three primitives in the presented methodology.

First, the targeted object is to be located on the vulnerable position as required in the primitive \emph{LP}. To implement it, the attacker can simply use the technique \emph{(A1) Object Spraying} to spray the memory with targeted pages in order to maximize the probability of corrupting the target. Or the attacker can also keep sending network packets to consume the available memory, then free the vulnerable position and pad the target into it precisely. In nature, this procedure is a variation of technique \emph{(A2) Forced Padding}. The privileged technique \emph{(A3) Induced Replacement} is not available for this remote unprivileged attack while the \emph{(A4) Try and Abort} has not been leveraged in such attacks. In short, locating the victim to desired position from the network origin can be achieved only by sending the network packets, which also means it need to exhaust a range of memory on the remote victim devices.

Performing \emph{RH} from network is very critical in the remote rowhammer attacks. The attacker has to realize both bypassing the cache and conducting hammering pattern only by sending and receiving network packets. Bypassing the cache is extremely difficult for a remote attacker who has no idea of victim's cache strategy for network packets. When the victim devices use RDAM  (\emph{Remote Direct Memory Access}) memory or \texttt{clflush} to handle network packets, the attacker might use the \emph{(Ba1) Specific Instructions} or \emph{(Ba3) Uncached Memory} to manipulate the network packets directly into the memory without caching. Even neither of two mechanisms are adopted in the victim devices, the attacker can still construct \emph{(Ba2) Cache Eviction Set} to bypass the cache by some special mechanism normally attached on the server, like Intel CAT~\cite{nethammer,aga2017good}. As for the choice of hammering pattern, since the attacker cannot gain address information of victim through the network, generally the technique \emph{(Bb1) Single-sided Hammering} or \emph{(Bb3) One-location Hammering} that require no  knowledge of address mapping is adopted~\cite{nethammer}. However, sometimes the attacker can leverage the remote server's own mechanisms to perform \emph{(Bb2) Double-sided Hammering}. For example, most servers tend to enable huge pages for storing mass data, so that attacker can make use of huge page to conduct double-sided hammering remotely~\cite{throwhammer}.

With the bit error injected by primitive \emph{RH}, the remote victim server is corrupted. To verify whether the bit flip is exploitable, the \emph{(C1) Directly Read} is available if the attack target is allowed to be checked, just simply sending request of accessing the target and then checking the content of feedback~\cite{throwhammer}. The technique \emph{(C2) Behaviour Judgement} for verification of injected bit flip is not that practical in the network rowhammer attacks. This is because the attacker can only communicate with the remote device through certain interfaces, and in general he cannot observe behaviors of the internal objects. The attacker can exploit
verified bit flip to conduct various applications, such as privileged escalation or DoS attack.

\subsection{Summary}
In this section, we summarize those presented rowhammer attacks according to our unified framework, discussing how to conduct an effective rowhammer attack from different origins to achieve intended implication.
Table~\ref{Table_existworks} lists all previously presented rowhammer attacks and displays primitives adopted in each attack, which are marked with $\checkmark$.
Note that there are still a number of
primitive combinations that have not been exploited in presented works. Future works can extract primitives from this table or even add new primitives to conduct novel rowhammer attacks.
{\rowcolors{7}{white}{gray!8}
\begin{table*}[!htb]
  \centering
  \caption{The analysis of existing rowhammer attacks based on the unified framework.}\label{Table_existworks}
  \scalebox{1}{
  \begin{tabular}{|c|c|c|c|c|c|c|c|c|c|c|c|c|c|c|c|c|c|c|c|c|}
  \hline
  \multirow{3}*{Attacks} & \multicolumn{4}{c|}{Attack Origin} & \multicolumn{4}{c|}{\multirow{2}*{Intended Implication}} & \multicolumn{12}{c|}{Methodology} \\
  \cline{2-5}\cline{10-21}
  & \multicolumn{2}{c|}{Local} & \multicolumn{2}{c|}{Remote} & \multicolumn{4}{c|}{} & \multicolumn{4}{c|}{LP} & \multicolumn{6}{c|}{RH} & \multicolumn{2}{c|}{EV} \\
  \cline{2-21}
  & \rotatebox{90}{\scriptsize Unprivileged Process} & \rotatebox{90}{\scriptsize Privileged Process} & \rotatebox{90}{\scriptsize Website} & \rotatebox{90}{\scriptsize Network} & \rotatebox{90}{\scriptsize Corrupt EPRO}
  & \rotatebox{90}{\scriptsize Corrupt EPUO} & \rotatebox{90}{\scriptsize Corrupt DPRO}
  & \rotatebox{90}{\scriptsize Corrupt DPUO} & \rotatebox{90}{\scriptsize (A1)Object Spraying} & \rotatebox{90}{\scriptsize (A2)Forced Padding} & \rotatebox{90}{\scriptsize (A3)Induced Replacement} & \rotatebox{90}{\scriptsize (A4)Try and Abort} &
  \rotatebox{90}{\scriptsize (Ba1)Specific Instructions} & \rotatebox{90}{\scriptsize (Ba2)Cache Eviction Set} & \rotatebox{90}{\scriptsize (Ba3)Uncached Memory} & \rotatebox{90}{\scriptsize (Bb1)Single-sided} &
  \rotatebox{90}{\scriptsize (Bb2)Double-sided} & \rotatebox{90}{\scriptsize (Bb3)One-location} & \rotatebox{90}{\scriptsize (C1)Directly Read} & \rotatebox{90}{\scriptsize (C2)Behaviour Judgement} \\
  \hline
  Flipping bits \cite{kimfirstpaper} & $\checkmark$ & & & & $\checkmark$ &  & & & $\checkmark$ & & & & & & & $\checkmark$ & $\checkmark$ & & $\checkmark$ & \\
  Gain kernel \cite{linuxkernel} & $\checkmark$ & & & & $\checkmark$ & & & $\checkmark$ & $\checkmark$ & & & & $\checkmark$ & & & $\checkmark$ & & & $\checkmark$ & $\checkmark$ \\
  New Approach \cite{newapproach}& $\checkmark$ & & & & $\checkmark$ & & & $\checkmark$ & $\checkmark$ & & & & $\checkmark$ & & & $\checkmark$ & & & $\checkmark$ & $\checkmark$ \\
  Rowhammer.js \cite{rowhammerjs} & $\checkmark$ & & $\checkmark$ & & & & & $\checkmark$ & $\checkmark$ & & & & & $\checkmark$ & & $\checkmark$ & & & & $\checkmark$ \\
  Dedup Est Machina \cite{dedupmachina}& & & $\checkmark$ & &    & $\checkmark$ & & &   & $\checkmark$ & & &    & $\checkmark$ & &  $\checkmark$  & & &    & $\checkmark$ \\
  One Bit Flips \cite{xiaoyuancloud}& & $\checkmark$ & & &  $\checkmark$  & & & &   & & $\checkmark$ & &  $\checkmark$  & & &    & $\checkmark$ & &  $\checkmark$  & \\
  Flip Feng Shui \cite{flipfengshui}& & $\checkmark$ & & &  $\checkmark$  & & & &   & & $\checkmark$ & &  $\checkmark$  & & &    & $\checkmark$ & &  $\checkmark$  & \\
  Drammer \cite{drammer}& $\checkmark$ & & & &    & & & $\checkmark$ &   & $\checkmark$ & & &    & & $\checkmark$ &    & $\checkmark$ & &    & $\checkmark$ \\
  Curious Case \cite{curious}& & $\checkmark$ & & &    & $\checkmark$ &  & &   & & & $\checkmark$ & $\checkmark$ & $\checkmark$ & &  $\checkmark$  & & &  & $\checkmark$ \\
  Good Go Bad \cite{aga2017good}& $\checkmark$ & & & &  $\checkmark$  & & & &  $\checkmark$ & & & &    & $\checkmark$ & &  $\checkmark$  & & &  $\checkmark$  & \\
  SGX-Bomb \cite{jang2017sgx}& $\checkmark$ & & & &    & $\checkmark$ & & &   & & & $\checkmark$ &  $\checkmark$  & & &  & $\checkmark$ & &    & $\checkmark$ \\
  Another Flip \cite{anotherflip}& $\checkmark$ & & & &    & & $\checkmark$ & &   & & & $\checkmark$ &  $\checkmark$  & & &    & & $\checkmark$ &  $\checkmark$  & \\
  Glitch \cite{glitch}& & & $\checkmark$ & &    & $\checkmark$ & & &   & $\checkmark$ & & &    & $\checkmark$ & &    & $\checkmark$ & &    & $\checkmark$ \\
  Throwhammer \cite{throwhammer}& & & & $\checkmark$ &    & & & $\checkmark$ &   & $\checkmark$ & & &    & & $\checkmark$ &    & $\checkmark$ & &    & $\checkmark$ \\
  RAMPAGE \cite{guardion} & $\checkmark$ & & & &    & & & $\checkmark$ &   & $\checkmark$ & & &    & & $\checkmark$ &    & $\checkmark$ & &    & $\checkmark$ \\
  Still hammerable \cite{stillhammer}& $\checkmark$ & & & &    & & & $\checkmark$ &   & $\checkmark$ & & &  $\checkmark$  & & &  $\checkmark$  & & &    & $\checkmark$ \\
  PFA \cite{pfa}& & $\checkmark$ & & &    & & $\checkmark$ & &   & & & $\checkmark$ &  $\checkmark$  & & &    & $\checkmark$ & &   $\checkmark$ & \\
  Nethammer \cite{nethammer}& & & & $\checkmark$ &    & $\checkmark$ & $\checkmark$ & $\checkmark$ &   & & & $\checkmark$ &   $\checkmark$ & $\checkmark$ & $\checkmark$ &    & & $\checkmark$ &  $\checkmark$  & $\checkmark$ \\
  \hline
\end{tabular}
  }
\end{table*}}

\section{countermeasures} \label{sec_counter}
With the rise of rowhammer attacks, a number of countermeasures are emerged. 
For instance, in the first paper exposing rowhammer~\cite{kimfirstpaper}, Kim et al. have proposed PARA (\emph{Probabilistic Adjacent Row Activation}) to prevent bit flips, and soon afterwards they propose two improved
countermeasures named as CRA (\emph{Counter-Based Row Activation}) and PRA (\emph{Probabilistic Row Activation})~\cite{kim2015architectural}.
In essence, all rowhammer countermeasures prevent the attack by blocking the attack process, \textit{i.e.}, the \emph{LP}, \emph{RH} and \emph{EV} primitives in the attack methodology.
Once one of the primitives is prevented, the attack will not work and cannot achieve the intended implication. In the following, we introduce existing rowhammer countermeasures and also propose our view on the defences.

\subsection{Prevent LP}
Locating victim object to vulnerable position is the precondition for an effective rowhammer attack. Hence there are
some countermeasures designed to prevent the \emph{Location preparation}. An empirical countermeasure is to mark all vulnerable
pages in the DRAM and forbid the system to use them, making the attacker incapable of locating the victim, like the B-CATT~\cite{touch}.
However, such countermeasure has been proved to be impractical and inherently insecure. The performance of system would be degraded as a large amount of memory is disabled.
Besides, experiment results in \cite{guardion} also show that the number of vulnerable locations increase during time,
it is hard to totally disable all of them. 
In most of previous rowhammer attacks~\cite{linuxkernel,rowhammerjs,drammer,guardion,glitch,throwhammer}, the attacker first needs to profile a large amount of memory for finding exploitable
bit flip, then further exhausts nearly the entire memory to position the victim on the vulnerable row. As a result, the locating always generates conspicuous memory footprints no matter whether
\emph{(A1) Object Spraying} or \emph{(A2) Forced Padding} is utilized. Hence, preventing the memory exhaustion is a possible defence to thwart
primitive \emph{LP}~\cite{rowhammerjs,drammer}. The memory allocator in the system should restrain exhausted memory usage and the OS also should kill the malicious process that consumes the entire memory.

For the \emph{(A3) Induced Replacement}, it normally adopts special mechanisms like para-virtualization~\cite{xiaoyuancloud} or memory deduplication~\cite{flipfengshui} to replace the victim with the forged object controlled by the attacker. Therefore, this technique is not universally applicable for all situations. Removing or disabling its dependent mechanisms is a possible way to prevent the locating of victim. The recently presented technique \emph{(A4) Try and Abort} is much stealthier
than other techniques. It does not exhaust the entire memory or rely on special mechanisms, while it only keeps trying different positions~\cite{anotherflip,pfa}. For now,
there are no formal papers presenting any effective defence against this locating technique. Perhaps monitoring the abnormal usage of cache can be feasible as it evicts the victim
from memory by increasing page cache pressure.

\subsection{Prevent RH}
As the most pivotal primitive in the rowhammer attack methodology, \emph{Rapid Hammering} is the major focus of nearly all countermeasures. To prevent the attacker from bypassing cache and performing
hammering, countermeasures are proposed at respective hardware and software levels. 

At the early time, countermeasures generally try to prevent \emph{Rapid Hammering} by modifying the hardware setting, eliminating the vulnerability fundamentally. As the primitive requires to hammer adjacent rows for
enough times in the refresh interval, one proposal is to double the refresh rate so that the hammering frequency is not enough for inducing bit flips~\cite{kimfirstpaper}. However, this defence would incur a high performance penalty and has been shown as ineffective in~\cite{anvil}. Besides that, some special mechanisms are also proposed, such as
ECC (\emph{Error Correcting Code}), PRA (\emph{Probabilistic Row Activation}) and TTR (\emph{Target Row Refresh}). ECC can detect and correct 1-bit error in the memory, however, it might meet the difficulty of coping with multiple bit flips~\cite{kimfirstpaper} by rowhammer. Nowadays, ECC is mainly deployed in the servers, and normal desktop and laptops may not support ECC mechanism. Besides,the latest work~\cite{ECCploit} successfully bypasses the ECC and can reliably flip bit on the ECC memory, which totally breaks such defence. TTR is adopted as an effective defence against rowhammer in the new standard LPDDR4~\cite{LPDDR4}. It refreshes adjacent rows if the targeted
row is accessed at a high frequency. The results in~\cite{drammer} show
that TTR cannot defeat rowhammer reliably as the attacker can still induce bit flips on a Google Pixel phone with 4GB LPDDR4 memory. Lipp et al.~\cite{nethammer}
also demonstrate that bit flips can be detected far away from the hammered rows, which means refreshing adjacent rows only is not enough for preventing bit flips.
PRA is a probability-based countermeasure enhanced from PARA. It simply allows memory controller to probabilistically open adjacent or non-adjacent rows, hoping to refresh vulnerable rows before bit flips occur. Such countermeasures can mitigate rowhammer attacks to some extent with less performance penalty, but its effectiveness needs further verification.

In recent years, more and more software-based countermeasures are proposed, as hardware-based ones cannot be directly attached to legacy systems. 
In terms of software, the simplest countermeasure is to remove \emph{(Ba1) Specific Instructions} used for bypassing the cache, such as disabling \texttt{clflush} in the NaCl
sandbox~\cite{linuxkernel}. Leveraging performance counter to detect abnormal frequent access is also a common countermeasure. Like the first concrete software-based
countermeasure ANVIL~\cite{anvil}, it utilizes CPU's performance counter to monitor the amount of last-level
cache misses, and marks the addresses if the amount exceeds a predetermined threshold. Once there are enough accesses to other rows in the same bank, the victim rows are
forced with an early refresh. ANVIL can effectively prevent rowhammer attack from injecting bit flip to the memory. However, the newly proposed
\emph{(Bb3) One-location Hammering} challenged this countermeasure, since it only accesses the same row and does not alert the ANVIL~\cite{anotherflip}. Recently, a
countermeasure is proposed to detect bit flips in~\cite{vig2018rapid}, which combines sliding window protocol and dynamic integrity hash
tree. Bit flips in the frequently accessed rows can be added to the hash tree for prompt detection. Such countermeasure mainly relies on detecting the injected bit flips rather than preventing them from the beginning of inducing.

\subsection{Prevent EV}
Let the attacker flip bit but prevent him from exploiting the flipped bit is also a feasible solution.
Even the attacker flips a bit in the victim, if he cannot exploit it to cross boundary between security domains, the flipped bit is not that meaning in terms of impacts. The core idea is to isolate physical memory to different domains and make sure that injected bit flips can only arise in the attacker's own memory region. G-CATT~\cite{touch} is an exemplary countermeasure to block primitive \emph{Exploit Verification}. It extends the memory allocator to physically isolate two domains, the kernel and the user space,
with gap rows. Even the attacker flips bits in the gap rows, he cannot induce bit flips in the kernel (\textit{e.g.}, page table) to achieve privilege escalation.
However, as to those shared objects in the system, the reliability of G-CATT is doubtful. Recent works show that CATT cannot manage to work. In~\cite{anotherflip}, one bit is flipped in the opcode of shared \emph{sudo} binary, and certain double-ownership kernel buffers like video buffers also allow the attacker to defeat such countermeasure~\cite{stillhammer}.

Indeed, the idea of preventing exploitation on the flipped bit is very novel and impressive, spawning a class of countermeasures. For instance, GuardION~\cite{guardion} is a similar countermeasure that intends to prevent exploiting of flipped bit on the
mobile devices. It use guard rows to isolate the DMA buffer, the main attack vector used for rowhammer attacks on mobile devices. It guarantees that the flipped bit only arises in the isolated DMA region, resulting in that there is no exploitable flips for attacker. Beyond that, the ALIS~\cite{throwhammer} proposed for defending Throwhammer also adopts guard rows to prevent exploiting of flipped bit. It is a new allocator to offer fine-grained memory isolation of network buffers, find guard rows at the DRAM address space and use them to isolate flipped bit from security
domains. The latest countermeasure ZebRAM~\cite{ZebRAM} even isolates all data rows with guard rows that absorb bit flips, which protects sensitive data from malicious exploiting.
Actually, preventing the primitive \emph{Exploit Verification} with guard rows is a promising direction for future design of countermeasure, since it is effective and practical.

\subsection{Summary}
In this section, we introduce existing rowhammer countermeasures based on which primitive in the methodology they try to prevent. Table~\ref{Table_counter} illustrates these countermeasures and shows their defensive effect.
\begin{table*}[!htb]
  \centering
  \caption{Countermeasures for rowhammer attacks.}\label{Table_counter}
  \scalebox{1}{
  \begin{tabular}{|c|c|p{9.5cm}|p{1.2cm}<{\centering}|}
  \hline
  Countermeasures & Affected Primitive & \centering Description & Reliability (For now) \\
  \hline
  B-CATT~\cite{touch} & LP & Disable vulnerable pages to prevent locating victim to such positions. & $\times$ \\
  \hline
  Detect memory footprint~\cite{rowhammerjs} & LP & Prevent memory exhaustion that is usually required for locating victim. & $\times$  \\
  \hline
  ECC~\cite{Hamming2014Error} & RH  & Detect and correct 1-bit error, preventing single bit flip in the memory. & $\times$ \\
  \hline
  TRR~\cite{LPDDR4} & RH  & Refresh adjacent rows if the targeted row is accessed at a high frequency. & $\times$ \\
  \hline
  PRA~\cite{kim2015architectural} & RH & Probabilistically open adjacent or non-adjacent rows to eliminate bit flips. & $\checkmark$ \\
  \hline
  ANVIL~\cite{anvil} & RH  & Use performance counter to monitor cache misses to detect rowhammer attack. & $\times$ \\
  \hline
  Disallow \emph{cluflush}~\cite{linuxkernel} & RH  & Disallow \emph{cluflush} to stop attacker from directly accessing the DRAM. & $\times$ \\
  \hline
  Double refresh rate~\cite{kimfirstpaper} & RH  & Double refresh rate of DRAM to mitigate change leak caused by hammering. & $\times$ \\
  \hline
  Detect with hash tree~\cite{vig2018rapid} & RH  & Combines sliding window protocol and integrity hash tree to detect bit flips. & $\checkmark$ \\
  \hline
  G-CATT~\cite{touch} & EV   & Physically isolate the memory of different system entities with a gap. & $\times$ \\
  \hline
  GuardION~\cite{guardion} & EV   & Isolate DMA buffer with guard rows to prevent mobile rowhammer attacks. & $\checkmark$ \\
  \hline
  ALIS~\cite{throwhammer} & EV  & Leverage a buffer allocator to offer fine-grained memory isolation. & $\checkmark$ \\
  \hline
  ZebRAM~\cite{ZebRAM} & EV  & Isolate all data rows with guard rows that absorb bit flips. & $\checkmark$ \\
  \hline
\end{tabular}
  }
\end{table*}

%
%
%
%
%
%
%
%
%
%
%

\section{Expressive Rowhammer Attacks} \label{sec_expressive}

By observing existing rowhammer works through the lens of our framework, we have identified a crucial limitation, expressiveness of attacks.
Rowhammer attacks rely on hardware faults present within the DRAM modules and hence position and/or patterns of possible bit flips are determined, posing a strict constraint on the \emph{LP} phase of the attack. As a result, existing rowhammer work typically uses simple, single bit flip to induce coarse changes in order to achieve the attack. This begs the question, how can we improve the expressiveness of rowhammer towards a meaningful and composable attack?

\begin{figure}[h]
  \centering
  \includegraphics[scale=0.18]{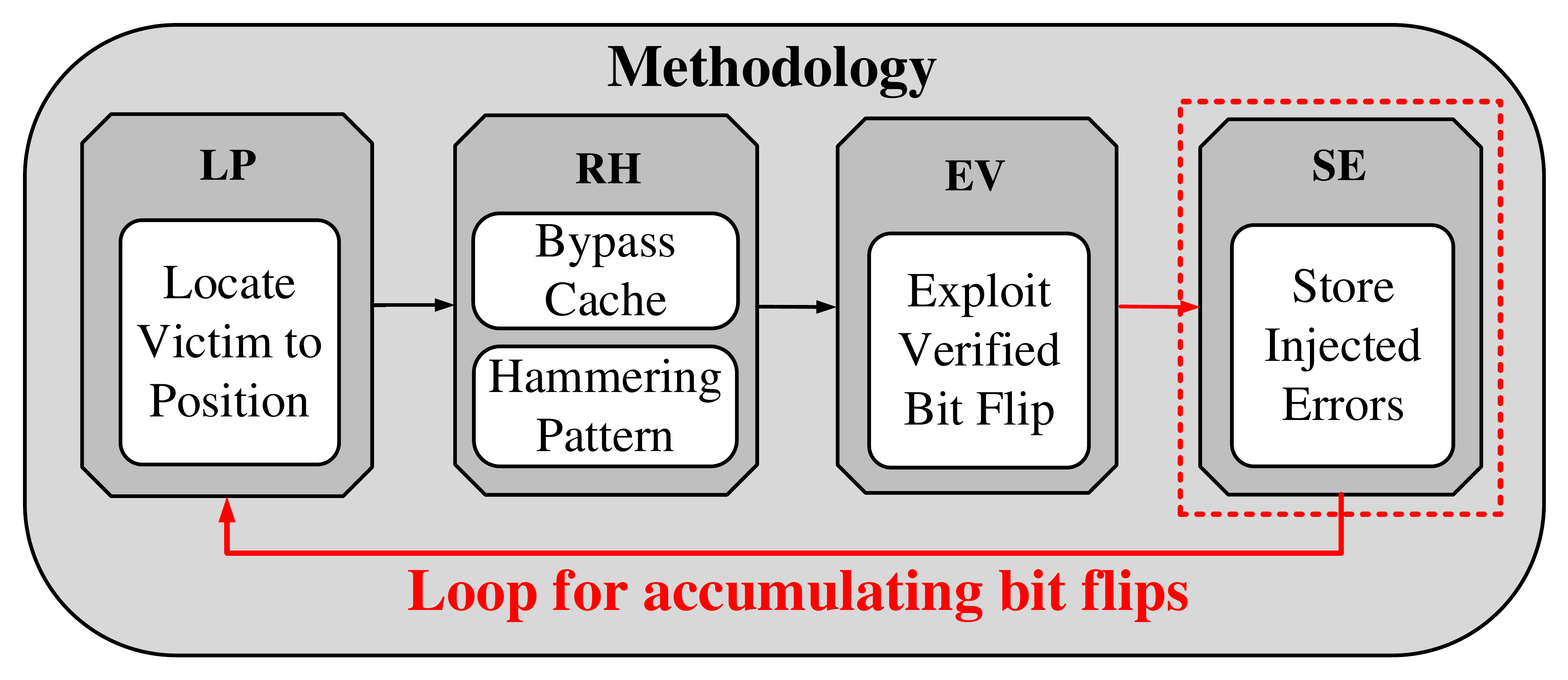}\\
  \caption{Methodology for expressive rowhammer attack.}\label{fig_express}
\end{figure}

To that end, we propose the addition of a new primitive \emph{SE (Store Error)} to the methodology as shown in Fig.~\ref{fig_express}. The addition of \emph{SE} allows the attacker to persist bit flips and by accumulating it, enables the composability of bit flips into complex bit patterns.
More concretely, the attacker first leverages \emph{LP} to steer the victim object towards the vulnerable position and
uses \emph{RH} to induce a bit flip in the victim. After verifying that the injected bit flip is exploitable based on the primitive \emph{EV}, the attacker can store the correct bit flip to the external storage such as the disk. At this point, a single bit error is persisted. By performing such loops, the attacker can accumulate multiple bit flips and precisely control their positions, achieving expressive changes to the victim object.

Intuitively, the threat of rowhammer attacks increases with the number of bit flips to be injected, so that such expressive multiple-bit error is much more threatening and can be applied to a wider range of scenarios than previous rowhammer attacks that generally only exploit single-bit error. Note that even the rowhammer bug is possible to induce multiple bit flips in a fixed memory region~\cite{kimfirstpaper,xiaoyuancloud}, the distribution of these multiple flips depends on the hardware property of the physical row and is always impossible to meet the demand of the actual attack. So, none of prior rowhammer works leverage multiple bit flips to achieve their attacks. In the following, we discuss how to implement the expressive rowhammer attack.

\subsection{Support Mechanism}
\emph{Store Error} is the key of conducting expressive rowhammer attack. It persists bit flips back to an external storage such that injected errors can be accumulated for expressive multiple-bit modification. However, its implementation requires the support from specific mechanisms, since corrupted victim object would be discarded in normal cases instead of storing back to the external storage.

Disk cache is a Linux kernel feature which corresponds to the memory cache. As a memory structure between the main memory and the disk, disk cache stores the data that are frequently accessed by the OS or user processes. Compared to the memory cache, disk cache can hold the data longer and have larger memory space. IO to the file is instead made to this structure in memory, which aims to reduce IO latency by utilizing free memory. Since the memory is technically free but is actually ``hijacked'' by the OS, the OS himself has to aggressively swap out the cache if programs request for memory. Further, to ensure cache coherence, the OS uses a dirty bit to mark that there is a difference between the file and the cache, and that an update to the file is required. As a result, once the OS flushes the modified disk cache, the dirty pages are stored back to the disk.

A challenge to address is that the modification by rowhammer is not mediated by the OS and will be ignored when flushing, so that we should leverage some mechanisms, such as certain OS utilities, to trigger the storing of errors caused by rowhammer. For the sensitive data stored in disk cache, the expressive rowhammer attack can be very powerful and practical. In fact, the page cache used in \cite{anotherflip} for their novel \emph{LP} method \emph{memory waylaying} is similar to the disk cache. However, they only leverage it to avoid causing large memory footprint. The usage for accumulating bit flips is not considered.

\subsection{Composing Expressive Rowhammer Attack}
\begin{figure}
  \centering
  \includegraphics[scale=0.23]{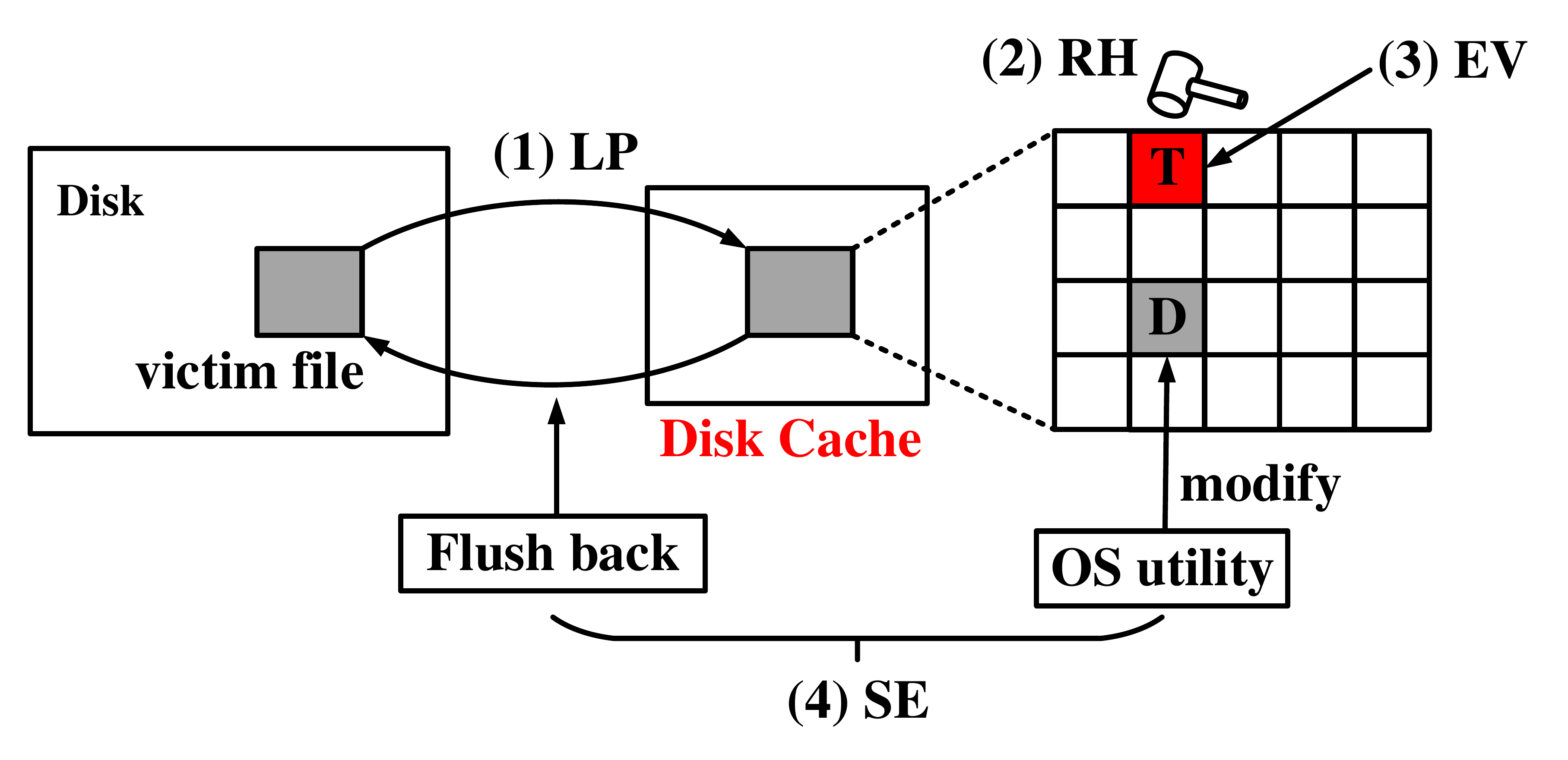}\\
  \caption{The concept of composing expressive rowhammer attack, $T$ denotes the victim and $D$ denotes the data used for legitimate modification.}\label{fig_mulattack}
\end{figure}
As an interesting mechanism that helps attacker to accumulate bit flips, we describe an expressive rowhammer attack via disk cache. Fig.~\ref{fig_mulattack} illustrates the conceptual composition of multiple bit flips, which contains four stages. 
\begin{itemize}
  \item Stage (1): \emph{Location Preparation (LP)}. Load the victim from the disk and locate it to vulnerable positions on the disk cache.
  \item Stage (2): \emph{Rapid Hammering (RH)}. Inject one bit flip into the victim $T$ by rapidly hammering adjacent rows.
  \item Stage (3): \emph{Exploit Verification (EV)}. Verify whether the injected error is one of exploitable bit flips.
  \item Stage (4): \emph{Store Error (SE)}. Modify data $D$ in the page containing the flipped bit with legitimate OS utilities, then flush the changed page to disk. This enables the injected bit flip to piggy-back on the flush and be persistently stored.
\end{itemize}

With the last stage \emph{SE} in Fig.~\ref{fig_mulattack}, the OS ignores modification caused by rowhammer when flushing the dirty disk cache pages. Therefore, we first leverage OS utilities to modify other data in the victim page to ensure the injected bit flip can be stored back together. However, the fulfillment of modification on data $D$ is not straightforward, which is dependent on the files that cached. We present an interesting technique in the following end-to-end attack. Ideally by repeating these four stages, a sequence of bit flips can be controlled towards an expressive meaning which relies on the context of the attack.

\subsection{An End-to-End Expressive Rowhammer Attack}
Combining the support of primitive \emph{Store Error} from disk cache and the original rowhammer attack methodology, we develop a powerful end-to-end expressive rowhammer attack that achieves privilege escalation by corrupting the file \emph{/etc/passwd}. It allows any normal user to change his user ID without the write permission.

\noindent\textbf{Target Identification.}
Many important and frequently used files are actually buffered in the disk cache, in order to prevent the performance decrease if those files are swapped in and out between the disk and the main memory. As a pivotal file in the Linux system, \emph{/etc/passwd} records many entries of user accounts, which is located in the disk cache once loaded.
The importance of \emph{/etc/passwd} makes it an ideal target, however, the direct attack on the sector of password is not easy. In fact, the Linux system stores the user password in an encrypted format into a different file, \textit{i.e.}, the \emph{/etc/shadow} file, which is marked as an ``x'' block and not readable for normal users.

A different sector called UID catches our attention. The UID determines the privilege level of a user~\cite{UID}. A privileged root, denoted as $\mathcal{R}$, has a UID of $0$. A normal user, denoted as $\mathcal{U}$, has a UID started from $1000$ in ASCII. Since the user with UID $1000$ is usually the root user of system, the UID $1001$ is a more interesting target to explore. Our expressive attack aims to modify it to $0$, so that the normal user can impersonate the \emph{root}.
Table~\ref{Tab-Binary} lists the corresponding binary representations of the UID sectors of the root $\mathcal{R}$ and the normal testing user $\mathcal{U}$. The modification from $0x31303031$ to $0x30303030$ requires a chain of two deterministic $1\rightarrow0$ bit flips in the \emph{/etc/passwd} file, which is impossible to accomplish using techniques from previous literatures.

\begin{table}[!htb]
  \centering
  \caption{Binary form of key data in the /etc/passwd.}\label{Tab-Binary}
  \scalebox{1}{
  \begin{tabular}{|l|l|l|}
  \hline
  role & ASCII code & Binary form  \\
  \hline
  root $\mathcal{R}$ & :x:\textcolor[rgb]{1.00,0.00,0.00}{0}: & 0x3a 0x78 0x3a \textcolor[rgb]{1.00,0.00,0.00}{0x30} 0x3a \\
  \hline
  tesing user $\mathcal{U}$ & :x:\textcolor[rgb]{1.00,0.00,0.00}{1001}: & 0x3a 0x78 0x3a \textcolor[rgb]{1.00,0.00,0.00}{0x31303031} 0x3a  \\
  \hline
\end{tabular}
  }
\end{table}

\noindent\textbf{Flipping a Single Bit.}
The \emph{/etc/passwd} file is well protected whose permission flag is set as ``-rw-r--r--''. Therefore normal users cannot write it except the root, which prevents the file from malicious modification. However, based on the aforementioned rowhammer attack methodology, we can craft effective attack method to alter one specific bit in \emph{/etc/passwd}. In fact, similar to the work in~\cite{anotherflip} that aims to modify opcode in the \emph{sudo} binary, the modification of \emph{/etc/passwd} file aims to corrupt a privileged binary file that is
shared by the user and kernel. As the \emph{/etc/passwd} file has only one copy in the memory, the spraying method like technique (A1) cannot work. We adopt the technique (A4) \emph{Try and Abort}, like the memory waylaying method in~\cite{anotherflip}), to locate the \emph{/etc/passwd} file on the vulnerable position. Then we select techniques from primitive \emph{RH}, such as (Ba1) \emph{Specific Instructions} and (Bb1) \emph{Single-sided Hammering}, to bypass the cache. We perform certain hammering pattern to inject bit flip to the victim, finally verify whether the flipped bit is exploitable by reading or judging victim's behavior.

\noindent\textbf{Chaining Multiple Bit Flips.}
After one round of single-bit flip, one of the required bits in the target UID is flipped. Note that the victim file is still saved in the
disk cache for the next access. He needs to persist the changes. To flush the flipped content to disk, we need the OS to treat the page
as dirty (or changed). This must be done in a way permitted by the OS. In fact, a normal user does have the need to update system
files such as {\em passwd}, to change his user information like address or the command shell. This can be achieved by the commands with \emph{suid}~\cite{suid}.
The attacker just needs to find an appropriate command with \emph{suid} to update his own credential in the \emph{/etc/passwd}
file, which will trigger a disk cache flush and cause the previously injected flip written back to the disk. There are many system commands
which can perform such an update. The most common one is \emph{passwd}. However, as mentioned earlier, the password is actually
stored in the \emph{/etc/shadow} file, so that it only flushes \emph{/etc/shadow} rather than \emph{/etc/passwd}. In our attack, the
command \emph{chsh}~\cite{chsh} is chosen for further exploitation. It allows normal users to modify their own shell information in the \emph{/etc/passwd} file, so that it can update the victim file without the root privilege.
After a single bit flip, the attacker can mark the \emph{/etc/passwd} file as dirty just by modifying his shell information in the file via the \emph{chsh} command. Then, he can run the \emph{sync} command which flushes all dirty files back to the disk without root privileges, so that the injected bit flip can be piggy-back on the flush and saved in the disk. As a result, the subsequent reloading of \emph{/etc/passwd} file will maintain the previous injected bit flip. In this way, multiple bit flips are chained together under the attacker's control, moving towards the expressiveness of a meaningful attack step by step.

\noindent\textbf{Root Privilege Escalation}
Our attack eventually updates the attacker's UID from $1001$ to $0000$. As the UID becomes $0$, the OS would misinterpret the normal
user $\mathcal{U}$ as \emph{root}. After logging in as $\mathcal{U}$, the attacker can perform root-level operations, such as changing permission flags of a file or modifying privileged system files. Hence, under our end-to-end attack, the normal user is equivalent
to the \emph{root} and the attacker achieves privilege escalation.

\section{Future Research Directions} \label{sec_future}

\subsection{Direction of Attacks}
With our proposed unified framework, novel rowhammer attack methods can be conducted by combining unused primitives or even adding new primitives.
In the following, we give a couple of directions for future research on the attacks.

\noindent\textbf{Expressive Rowhammer Attack.}
As discussed in section~\ref{sec_expressive}, compared to previous attacks that only exploit one single bit error, expressive rowhammer attack is much more powerful and also has wider application scenarios.
Hence, promoting the expressivity of rowhammer attacks is a promising direction for future research, which can greatly increase the threat of attacks. In our method, we adopt the disk cache as the support mechanism
for storing injected error, we believe there are more unknown mechanisms can be used for achieving such expressive attacks.

\noindent\textbf{Network Rowhammer Attack.}
Originating rowhammer attacks from network is the latest trend, and perhaps it could become the mainstream of future research with the rapid development of network.
As the network rowhammer attack is enough stealthy and practical, it is able to wreak havoc on the remote server without attracting attention, which exposes the server to the great threat.
It is necessary to analyze such novel attack and enable proactive prevention.

\noindent\textbf{Mobile Rowhammer Attack.}
Nowadays, mobile devices go deeper into people's lives than computers. As a result, mobile rowhammer attack is also a possible direction for future research. Actually, there are already a number of attacks aiming
to corrupt sensitive data or even escalate privilege from the mobile devices being proposed. However, as the upgrade of mobile devices is generally much faster, the corresponding attack techniques also
need to be updated.

\subsection{Direction of Countermeasures}
The rising of rowhammer attacks accelerate the research on the countermeasures, various defences are presented in order to effectively defeat the attack. However, up to now, there are no countermeasures can absolutely
defend against all rowhammer attacks. Hence, crafting more effective countermeasures is very important in the future research.

\noindent\textbf{Hardware Solutions.}
The rowhammer bug is the inherent vulnerability in the modern DRAM modules which have high cell density. Therefore, to solve this problem fundamentally, we need to improve the structure of hardware, eradicating
the disturbance errors in the memory without affecting its performance. It is a possible direction for designing countermeasures, even it may be hard to be achieved and cannot be applied on the legacy systems.

\noindent\textbf{Probabilistically Refresh.}
Refreshing the victim row before it is flipped can effectively prevent the rowhammer attacks. But in practical attacks, the judgement of victim row is difficult since the pattern or frequency of
hammering has no strict standards, resulting the countermeasure is insufficient for defending attacks. Hence, probabilistically refreshing rows may be a better solution. It randomly
refresh all rows in the memory and doesn't mark victim rows, which has a high probability of interrupting the process of attack.

\noindent\textbf{Isolate with Gap.}
As the most popular direction of designing countermeasures, isolating different domains' memory with guard rows shows strong effectiveness on defending rowhammer attacks. It trades off between security and performance,
sacrificing part of rows to gap the sensitive data from the attacker. Such countermeasure is able to prevent the attacker exploiting his injected bit error, and can be achieved with only software, so that it can be
applied on the legacy systems. Future works can keep going in this direction and try to propose more software-based countermeasures.

\section{Conclusion} \label{sec_conclusion}
Since the disclosure of rowhammer, a large number of attacks leveraging it are presented, which break the memory isolation between processes and make a great threat to the system security.
We propose a unified reference framework to systematically analyze rowhammer attacks, discussing about the attack origin and the intended implication, and further providing the methodology for conducting
effective rowhammer attacks.
Using the proposed framework, we analyze existing attacks and corresponding countermeasures, extracting primitives used in them.
Further, we also demonstrate the usefulness of such a framework by highlighting how researchers is able to find new combination of existing primitives or even add new primitives to achieve novel rowhammer attacks, such as the proposed expressive rowhammer attack using the disk cache.
Finally, we outlines possible future directions of rowhammer attacks.

\nocite{*}

\bibliographystyle{plain}
\bibliography{ref}

\end{document}